\documentclass{IEEEtran} 

\usepackage{graphics} 
\usepackage{amsmath} 
\usepackage{amssymb}  
\usepackage{amsthm} 
\usepackage{tikz} 
\usepackage{subfigure}
\usepackage{enumerate} 
\usepackage{amstext}  
\usepackage{mathrsfs} 
\usepackage[abs]{overpic}

\DeclareMathOperator*{\argmin}{arg\,min} 
\DeclareMathOperator*{\argmax}{arg\,max} 

\newtheorem{theorem}{Theorem} 
\newtheorem{lemma}{Lemma} 
\newtheorem{definition}{Definition} 
\newtheorem{remark}{Remark} 
\newtheorem{corollary}{Corollary} 
 
\newtheorem{proposition}{Proposition} 
\newtheorem{example}{Example}

\def\BibTeX{{\rm B\kern-.05em{\sc i\kern-.025em b}\kern-.08em
    T\kern-.1667em\lower.7ex\hbox{E}\kern-.125emX}}

\begin{document}
\title{A Cascade of Systems and the Product of Their $\theta$-Symmetric Scaled Relative Graphs}

\author{Xiaokan Yang, 
Ding Zhang, 
Wei Chen, 
and Li Qiu, \IEEEmembership{Fellow, IEEE}
\thanks{This work was supported in part by the National Natural Science Foundation of China under grants 62073003 and 72131001, by the Natural Science Foundation of Guangdong Province under grants 2024A1515011630 and by the Research Grants Council of Hong Kong under the General Research Fund No. 16206324.}
\thanks{Xiaokan Yang is with the School of Advanced Manufacturing and Robotics,  Peking University, Beijing 100871, China  (e-mail: yxkan21@stu.pku.edu.cn).}
\thanks{Ding Zhang is an independent scholar (e-mail: ding.zhang@connect.ust.hk).}
\thanks{Wei Chen is with the School of Advanced Manufacturing and Robotics and the State Key Laboratory for Turbulence and Complex Systems, Peking University, Beijing 100871, China  (e-mail: w.chen@pku.edu.cn).}
\thanks{Li Qiu is with School of Science and Engineering, The Chinese University of Hong Kong, Shenzhen, Guangdong, China  (e-mail: qiuli@cuhk.edu.cn).}
}

\maketitle

\begin{abstract}
In this paper, we utilize a variant of the scaled relative graph (SRG), referred to as the $\theta$-symmetric SRG, to develop a graphical stability criterion for the feedback interconnection of a cascade of systems. A crucial submultiplicative property of $\theta$-symmetric SRG is established, enabling it to handle cyclic interconnections for which conventional graph separation methods are not applicable. By integrating both gain and refined phase information, the $\theta$-symmetric SRG provides a unified graphical characterization of the system, which better captures system properties and yields less conservative results. In the scalar case, the $\theta$-symmetric SRG can be reduced exactly to the scalar itself, whereas the standard SRG appears to be a conjugate pair. Consequently, the frequency-wise $\theta$-symmetric SRG is more suitable than the standard SRG as a multi-input multi-output extension of the classical Nyquist plot. Illustrative examples are included to demonstrate the effectiveness of the $\theta$-symmetric SRG. 
\end{abstract}

\begin{IEEEkeywords}
$\theta$-symmetric scaled relative graph, Cascaded systems, Stability analysis, Nyquist plot 
\end{IEEEkeywords}

\section{Introduction}              \label{section:introduction} 

Graphical analysis is a fundamental method widely used in classical control theory and applications. In particular, the Nyquist plot of single-input single-output (SISO) linear time-invariant (LTI) systems, integrating both gain and phase information, provides crucial insights into the stability, robustness and performance of closed-loop systems \cite{Qiu2009IntroductionFeedbackSystems,Liu2016Robust_Control}. 

For multi-input multi-output (MIMO) LTI systems, the eigenloci---trajectories of eigenvalues of the transfer matrix along the Nyquist contour---play a fundamental role in the stability analysis. The generalized Nyquist criterion provides a necessary and sufficient condition for closed-loop stability in terms of eigenloci \cite{MacFarlane1977GeneralizedNyquist,Desoer1980Generalized}. However, unlike the SISO case where the Nyquist plot of the loop transfer function can be built directly from those of its components, the eigenloci of the MIMO return ratio bear no explicit connection to the eigenloci of each component. In other words, the closed-loop eigenloci cannot be inferred from the behavior of individual components. Moreover, for large-dimensional transfer matrices, plotting the eigenloci becomes numerically challenging. 

A classical method to overcome the aforementioned issues is to employ graphical over-approximation of the eigenloci, which is expected to enclose all eigenloci as tightly as possible while maintaining desirable analytical properties. Specifically, the well-known gain-based $\mathcal{H}_\infty$ theory offers a graphical method by constraining the gain within a disk \cite{Zhou1996Robust,Liu2016Robust_Control}. The small gain theorem states that if, at every frequency, the product of the maximum singular values of all loop components is less than one, then the closed-loop system is stable. In parallel with the gain, considerable effort has been devoted to developing the phase counterpart recently. Some notable developments include sectorial phases \cite{Chen2024Phase,Mao2022Phases,Wang2024First_five_year,Tits1999Robustness}, defined through the numerical ranges of matrices; singular angles \cite{Chen2025SingularAngle,Wielandt1967Topics,Gustafson1968Angle,Krein1969AngularLocalization,Gustafson1994Antieigenvalues}, which are based on the angles between vectors; and segmental phases \cite{Chen2025Cyclic}, which refine and extend the notion of singular angles. Additionally, the theories of positive real systems \cite{Anderson1973Network} and negative imaginary systems \cite{Petersen2010Feedback_control_NI} are closely related to phase, as both incorporate qualitative phase information of the systems. 

Combining gain and phase together offers a more uniform framework than treating them separately. In particular, the scaled relative graph (SRG), originally proposed in the optimization literature \cite{Hannah2016ScaledRelativeGraph,Ryu2022SRG} and subsequently brought into control community by \cite{Pates2021Scaled,Chaffey2022Rolledoff,Chaffey2023Graphical}, has attracted considerable attention and inspired many follow-up works \cite{vandenEijnden2025Phase,Chen2025Graphical,Krebbekx2025Graphical,deGroot2025DissipativitySRG,Baron2025MixedGainPhase,Baron2025StabilityLTI_SRG}. At the same time, however, it is important to recognize certain limitations of the current formulation. For example, the SRG definition introduces conjugate terms, and thus the SRG of a complex scalar matrix appears as a conjugate pair. Besides adding conservatism, this observation suggests that the SRG definition may not exactly reduce to the scalar case. This is a limitation of the standard SRG definition, which stems from a more fundamental problem: how to appropriately define the angle between two complex vectors? 

Motivated by this, we begin by proposing a more general definition of the angle between two complex vectors that encompasses the classical definition as a special case. Building on this, we then develop a variant of the SRG, referred to as the $\theta$-symmetric SRG, which coincides with the rotated SRG concept originally proposed in \cite{Zhang2025DW_Shell}.  

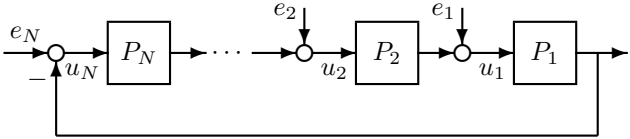
\begin{figure}[htb] 
    \setlength{\unitlength}{1mm}
		\begin{center}
			\begin{picture}(84,20)
				\thicklines 
                \put(0,14){\vector(1,0){6}}
                \put(7,14){\circle{2}} 
                \put(8,14){\vector(1,0){6}} 
                \put(14,10){\framebox(8,8){$P_N$}} 
                \put(22,14){\vector(1,0){5}} 
                \put(30,14){\makebox(0,0){$\cdots$}} 
                \put(33,14){\vector(1,0){6}} 
                \put(40,14){\circle{2}} 
                \put(40,20){\vector(0,-1){5}} 
                \put(41,14){\vector(1,0){6}} 
                \put(47,10){\framebox(8,8){$P_2$}} 
                \put(55,14){\vector(1,0){5}} 
                \put(61,14){\circle{2}} 
                \put(61,20){\vector(0,-1){5}} 
                \put(62,14){\vector(1,0){6}} 
                \put(68,10){\framebox(8,8){$P_1$}} 
                \put(76,14){\vector(1,0){7}} 
                \put(79,14){\line(0,-1){11}} 
                \put(79,3){\line(-1,0){72}} 
                \put(7,3){\vector(0,1){10}} 
                \put(3,9){\makebox(3,3){$-$}} 
                \put(1.5,15){\makebox(3,3){$e_N$}} 
                \put(9,10){\makebox(3,3){$u_N$}} 
                \put(36,18){\makebox(3,3){$e_2$}} 
                \put(42.5,10){\makebox(3,3){$u_2$}} 
                \put(57,18){\makebox(3,3){$e_1$}} 
                \put(63.5,10){\makebox(3,3){$u_1$}} 
			\end{picture}
			\vspace{-0.5em} 
                \caption{Feedback interconnection of a cascade of systems. }
			\label{fig:cyclic_feedback_system} 
		\end{center} 
\end{figure}

In this paper, we focus on the feedback interconnection of a cascade of systems as shown in Fig. \ref{fig:cyclic_feedback_system}, which arises in various application domains such as biochemical networks \cite{Tyson1978Dynamics,Hori2011CyclicGene} and large-scale systems  \cite{Arcak2006DiagonalStability,Sontag2006SecantCondition,Jiang2008Generalization,Pates2023Generalisation}. We develop graphical stability criteria for such cyclic interconnections, and further demonstrate that the $\theta$-symmetric SRG framework subsumes several existing results as special cases and yields less conservative stability conditions. 
For the case of two interconnected components, the graph separation methods in \cite{Zhang2025DW_Shell,Liang2024Feedback,Li2008Eigenvalues} suffice to provide a clear stability analysis. In contrast, the cyclic interconnection of cascaded systems renders these separation methods no longer applicable. In this work, we establish a crucial submultiplicative property of $\theta$-symmetric SRG, thereby making the $\theta$-symmetric SRG framework suitable for stability analysis of cyclic interconnections. This advancement also lays a foundation for potential extensions of the $\theta$-symmetric SRG framework to more general networked systems beyond cyclic structures. 

The rest of the paper is organized as follows. Section \ref{section:preliminaries} provides a brief review of the background and basic aspects of the SRG. In Section \ref{section:matrix_theta_SRG}, we first propose a general definition of the angle between complex vectors and then use it to introduce the notion of $\theta$-symmetric SRG along with its properties. The $\theta$-symmetric SRG of the cascaded product of matrices is analyzed, and several results in the existing literature can be recovered via over-approximations of the $\theta$-symmetric SRG. Section \ref{section:stability_systems} is dedicated to a graphical stability criterion for the feedback interconnection of a cascade of systems based on the $\theta$-symmetric SRG. Two examples are included in Section \ref{section:examples} to illustrate the effectiveness of $\theta$-symmetric SRG. Section \ref{section:conclusion} concludes the paper. 

The notation used in this paper is mostly standard. Let $\mathbb{R}$ and $\mathbb{C}$ be the set of real and complex scalars. The transpose, conjugate and conjugate transpose are denoted by $(\cdot)^T, \overline{(\cdot)}$ and $(\cdot)^H$, respectively. The inner product between two vectors $x,y\in\mathbb{C}^{n}$ is denoted by $\langle x, y\rangle = x^Hy$. The Euclidean norm of a vector $x \in \mathbb{C}^{n}$ is denoted by $\| x \|=\sqrt{\langle x,x\rangle}$. The spectrum of a matrix $C \in \mathbb{C}^{n \times n}$ is denoted by $\Lambda(C)$. Denote the real part and imaginary part of a matrix $C$ as $\mathrm{Re} \,C = \frac{1}{2}(C+C^H)$ and $\mathrm{Im}\, C = \frac{1}{2j}(C - C^H)$, respectively. The largest and smallest singular values of $C$ are denoted by $\overline{\sigma}(C)$ and $\underline{\sigma}(C)$, respectively. Let $\mathcal{RH}_\infty^{m\times m}$ denote the set of $m \times m$ real rational proper stable transfer matrices.

\section{Preliminaries on SRG}             \label{section:preliminaries} 

\subsection{Definition of SRG}      
\label{subsection:SRG_definition} 

We first review the classical definition of angles between vectors. For given vectors $x,y\in\mathbb{C}^{n}$, the angle $\angle (x,y) \in [0,\pi]$ between $x$ and $y$ is defined by \cite{Wielandt1967Topics,Gustafson1997NumericalRange} 
\begin{equation}                        \label{eq:vector_angle}
    \angle(x,y) = \arccos \frac{\mathrm{Re}\,\langle x,y \rangle}{\| x \| \|y\|}, \quad \text{ if } \  x \neq 0, y \neq 0 ,    
\end{equation} and $\angle (x,y) = 0$ if either $x=0$ or $y=0$. This notion of angle between vectors satisfies the triangle inequality, which is important in the feedback stability analysis. 

\begin{lemma}[\cite{Gustafson1997NumericalRange,Ryu2022SRG}]             \label{lem:triangle_inequality_signals}
    Let $x,y,z\in \mathbb{C}^{n}$ be nonzero vectors, then 
    \begin{equation*}
        |\angle(x,y) - \angle(y,z)| \leq \angle(x,z) \leq \angle(x,y) + \angle(y,z) . 
    \end{equation*} 
\end{lemma}  

Now we give a brief introduction of SRG. The notion of the SRG was first defined and presented in \cite{Hannah2016ScaledRelativeGraph}, where it serves as a new geometric tool to analyze contractive and non-expansive fixed-point iterations. It is defined for operators on real Hilbert spaces. Specializing to the finite-dimensional matrix case, the SRG of a real matrix $R \in \mathbb{R}^{n\times n}$ is defined as \cite{Huang2019SRG_Normal} 
\begin{equation*}
    \mathcal{G}(R) = \left\{  \frac{\|Rx\|}{\| x \|} \exp\{ \pm j \angle (x,Rx) \}: 0\neq x \in \mathbb{R}^{n} \right\}, 
\end{equation*}
which is a subset of $\mathbb{C}$ and symmetric with respect to the real axis. The SRG defined in this way for a real matrix is reasonable and natural, in view that the spectrum of real matrices is symmetric with respect to the real axis. 

When the concept of SRG was introduced to the control theory in \cite{Pates2021Scaled,Chaffey2022Rolledoff,Chaffey2023Graphical}, it was adapted for operators on complex Hilbert spaces. The SRG of a complex matrix $C \in \mathbb{C}^{n\times n}$ is defined in a similar manner:  
\begin{equation}                    \label{eq:SRG_matrix} 
    \mathrm{SRG}(C) = \left\{  \frac{\|Cx \|}{\| x\|} \exp\{ {\pm j\angle(x,Cx)} \}: 0 \neq x \in \mathbb{C}^n \right\} ,  
\end{equation}
which is also symmetric with respect to the real axis. In the case of complex matrices, such symmetry is undesirable, as the real axis has no privileged status---unlike in the real matrix setting. 
The inclusion of the conjugate terms in \eqref{eq:SRG_matrix} makes $\mathrm{SRG}(C)$ coincide with $\mathrm{SRG}(\overline{C})$ and leads to the SRG of a complex scalar appearing as a conjugate pair. An illustrated example is shown as follows. 

\begin{example}     
    Consider the following cases: 
    \begin{enumerate}[(i)]
        \item Let $ C_1 \!  = \! j,   C_2\!  = \! -j,   C_3 \! = \!  
    \begin{bmatrix}    j & 0 \\ 0  & -j   \end{bmatrix} . $ 
    Then 
    \begin{equation*}
        \mathrm{SRG}(C_1)=\mathrm{SRG}(C_2)=\mathrm{SRG}(C_3)=\{ j, -j \} , 
    \end{equation*} 
    which is a conjugate pair. 
    \item Let $C = \begin{bmatrix}
        j & 0 \\ 0 & 1 
    \end{bmatrix}.$ Then 
    \begin{equation*}
        \mathrm{SRG}(C) = \mathrm{SRG}(\overline{C}) = \left\{ e^{j\phi} \!\in\! \mathbb{C}: \phi \!\in\! \left[-\frac{\pi}{2}, \frac{\pi}{2}\right] \right\}. 
    \end{equation*} 
    \end{enumerate} 
\end{example}

\subsection{Properties of SRG} 

Some basic properties of SRG are listed as follows  \cite{Ryu2022SRG,Chaffey2023Graphical,Chaffey2022Rolledoff,Pates2021Scaled,Huang2019SRG_Normal}.  

\begin{lemma}      \label{lem:property_SRG}
    Let matrix $C \in \mathbb{C}^{n\times n}$ be given, then the following statements are true: 
    \begin{enumerate}[(i)]
        \item Spectrum containment $\Lambda(C) \subset \mathrm{SRG}(C)$ holds. 
        \item For $0\neq \mu \in \mathbb{R}$, there hold  
        \begin{align*}
            \mathrm{SRG} (\mu C) & = \mu \mathrm{SRG}(C) , \\ 
            \mathrm{SRG}(C+\mu I) & = \mathrm{SRG}(C) + \mu . 
        \end{align*} 
        \item If $C$ is nonsingular, then 
        \begin{equation*}
            \mathrm{SRG}(C^{-1})  = (\mathrm{SRG}(C))^{-1} . 
        \end{equation*} 
        To clarify, $(\mathrm{SRG}(C))^{-1} = \left\{  z^{-1} \in \mathbb{C}: z \in \mathrm{SRG}(C) \right\}$. 
    \end{enumerate} 
\end{lemma}

The SRG of a matrix set $\mathcal{C} \subset \mathbb{C}^{n \times n}$ is defined as 
\begin{equation*}
    \mathrm{SRG}(\mathcal{C}) = \bigcup_{C \in \mathcal{C}} \mathrm{SRG}(C). 
\end{equation*}
Define the line segment between $z_1,z_2\in\mathbb{C}$ as $[z_1,z_2] = \{\mu z_1 + (1-\mu) z_2 : \mu \in [0,1]\}$. Then $\mathcal{C}$ is said to satisfy the chord property if, for all $z \in \mathrm{SRG}(\mathcal{C})$, there holds $[z,\overline{z}] \subset \mathrm{SRG}(\mathcal{C}) $.  
Define the arcs between $z$ and $\overline{z}$ as 
\begin{align*}
    \mathrm{Arc}^+(z,\overline{z}) & = \{ |z| e^{j\mu\angle z} : \mu \in [-1,1] \},  \\ 
    \mathrm{Arc}^-(z,\overline{z}) & = - \mathrm{Arc}^+(-z,-\overline{z}) . 
\end{align*}
Then $\mathcal{C}$ is said to satisfy the arc property if either $\mathrm{Arc}^+(z,\overline{z}) \subset \mathrm{SRG}(\mathcal{C}) $ for all $z \in \mathrm{SRG} (\mathcal{C})$, or $\mathrm{Arc}^-(z,\overline{z}) \subset \mathrm{SRG}(\mathcal{C}) $ for all $z \in \mathrm{SRG} (\mathcal{C})$.  

\begin{lemma}[\cite{Ryu2022SRG,Chaffey2023Graphical}]               \label{lem:SRG_chord_arc_property} 
    Let matrix sets $\mathcal{A},\mathcal{B} \subset \mathbb{C}^{n\times n}$ be given, then 
    \begin{enumerate}[(i)]
        \item if either $\mathcal{A}$ or $\mathcal{B}$ satisfies the chord property, then 
        \begin{equation*}
            \mathrm{SRG}(\mathcal{A} + \mathcal{B}) \subset \mathrm{SRG}(\mathcal{A}) + \mathrm{SRG}(\mathcal{B}) ; 
        \end{equation*} 
        \item if either $\mathcal{A}$ or $\mathcal{B}$ satisfies the arc property, then 
        \begin{equation*}
            \mathrm{SRG}(\mathcal{A} \mathcal{B}) \subset \mathrm{SRG}(\mathcal{A}) \mathrm{SRG}(\mathcal{B}). 
        \end{equation*}
    \end{enumerate}
\end{lemma}

Indeed, many typical uncertainty sets, including norm-bounded matrices and positive real matrices, naturally satisfy both chord and arc properties. This indicates that the chord and arc properties of matrix sets above are not ad hoc but rather reflect some general principles underlying the uncertainty sets.

\section{$\theta$-Symmetric SRG of Complex Matrices}          \label{section:matrix_theta_SRG} 
\subsection{Definition and Properties}        
\label{subsection:definition_theta_SRG_matrix} 

As discussed in Section \ref{subsection:SRG_definition}, the current SRG formulation exhibits certain limitations that appear to stem from a more fundamental issue, namely the definition of the angle between complex vectors. The definition in \eqref{eq:vector_angle} is classical \cite{Gustafson1997NumericalRange} and often taken for granted, yet it relies exclusively on the real part of the complex inner product. Such a convention deserves to be questioned. In complex spaces, the real part is not intrinsically special, and using it to define angles effectively privileges the real axis as a reference direction---an arbitrary choice lacking geometric justification. For instance, consider any nonzero $x \in \mathbb{C}^n$ and let $y = jx$. According to \eqref{eq:vector_angle}, the angle between $x$ and $y$ is $\frac{\pi}{2}$, whereas using the imaginary part instead yields the angle is $0$. 

Motivated by this, we begin with a more general definition of the angle between complex vectors. Then we adopt this definition to formulate the $\theta$-symmetric SRG and derive its properties and implications. 

For given vectors $x, y \in \mathbb{C}^{n}$ and $\theta \in \mathbb{R}$, define $\angle_\theta(x,y) \in [0,\pi]$ as 
\begin{equation}                \label{eq:vector_angle_phase_related}
    \angle_\theta(x,y) = \arccos \frac{\mathrm{Re}\langle x, e^{-j\theta} y \rangle}{\|x\|\|y\|} \quad \text{ if } \  x \neq 0, y \neq 0 ,    
\end{equation} 
and $\angle_\theta (x,y) = 0$ if either $x=0$ or $y=0$. A related study can be found in \cite{Paul2015Computation_theta_antieigenvalues}.  Based on \eqref{eq:vector_angle_phase_related}, the $\theta$-symmetric SRG of a matrix $C \in \mathbb{C}^{n\times n}$ is defined as 
\begin{equation}                \label{eq:definition_PS_SRG_matrix}
    \mathrm{SRG}_\theta(C) \! = \! \left\{ \! 
        \frac{\|  Cx \| }{\| x \|} \exp\{{j(\theta \! \pm \! \angle_\theta(x, Cx)}\} \! :   0 \! \neq \! x \! \in \! \mathbb{C}^n \! 
    \right\} . 
\end{equation} 
In view of \eqref{eq:SRG_matrix}, an explicit connection between $\theta$-symmetric SRG and the standard SRG is 
\begin{equation}                            \label{eq:definition_theta_SRG_matrix} 
    \mathrm{SRG}_\theta (C) = e^{j\theta} \mathrm{SRG}(e^{-j\theta} C) , 
\end{equation} 
which coincides with the rotated SRG concept introduced in \cite{Zhang2025DW_Shell}. Clearly, when $\theta = 0$,  $\mathrm{SRG}_0(C)$ reduces to the standard $\mathrm{SRG}(C)$ given in \eqref{eq:SRG_matrix}. For any $\theta\in\mathbb{R}$, $\mathrm{SRG}_\theta(C)$ is a subset of $\mathbb{C}$ and is symmetric with respect to the line $\{ \mu e^{j\theta} : \mu \in \mathbb{R} \}$. This property motivates the terminology ``$\theta$-symmetric SRG" that we adopt here. 

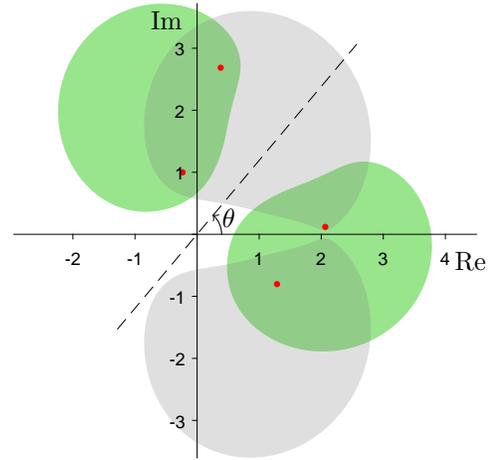
\begin{figure}[htb]
    \centering
    \begin{overpic}[width=0.7\linewidth]{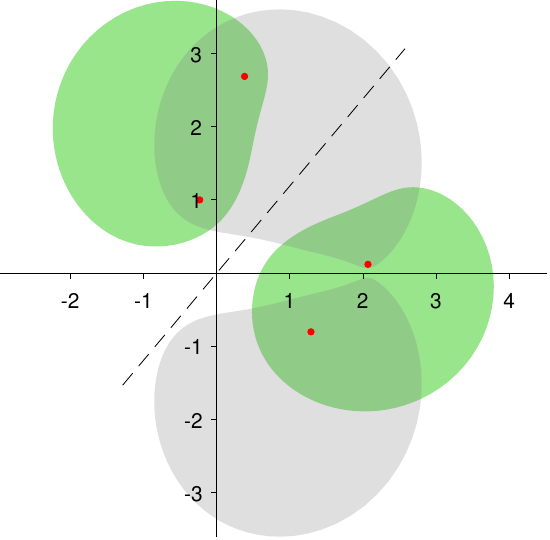}
        \put(79,88){$\theta$}   
        \put(167,72){$\mathrm{Re}$} 
        \put(52,163){$\mathrm{Im}$} 
        \begin{tikzpicture}[overlay] 
            \draw[->] (2.77,3) arc(0:48.5:0.35);  
        \end{tikzpicture} 
    \end{overpic}
    \caption{An illustration of $\mathrm{SRG}(C)$ (gray region) and $\mathrm{SRG}_\theta(C)$ (green region) with $\theta = 50^\circ$, where the red points are the eigenvalues of $C$. }
    \label{fig:SRG_theta_SRG} 
\end{figure}

\begin{example}
    Consider a matrix 
    \begin{equation*}
        C = \begin{bmatrix}
            1+2j & 0 & 2 & 0 \\ 
            0 & 1+j & -1 & 1 \\ 
            -j & 1 & 1.5 & 0 \\ 
            2 & -j & 1 & 0 
        \end{bmatrix}. 
    \end{equation*} Let $\theta = 50^\circ$, then an illustration of $\mathrm{SRG}(C)$ and $\mathrm{SRG}_\theta(C)$ is shown in Fig. \ref{fig:SRG_theta_SRG}. 
\end{example}

With $\theta$ serving as a parameter, the $\theta$-symmetric SRG refines phase information compared to the standard SRG. For given $\theta\in \mathbb{R}$, the phase spread of the $\theta$-symmetric SRG is defined as 
\begin{equation}             \label{eq:phase_spread}
    \Gamma_\theta(C) = \sup_{x \in \mathbb{C}^n} \,  \angle_\theta (x,Cx) . 
\end{equation}
Based on \eqref{eq:vector_angle_phase_related}, define the optimal $\theta^\star \in \mathbb{R}$ as 
\begin{equation}                        \label{eq:optimal_theta}  
    \theta^\star = \argmin_{\theta\in\mathbb{R}} \,  \Gamma_\theta(C) , 
\end{equation}
to minimize the phase spread of the corresponding $\theta$-symmetric SRG. The benefit of this optimization can be illustrated by the following example. 

\begin{example}             \label{examp:for_optimal_theta_SRG} 
     Consider the following sets of matrices: 
    \begin{enumerate}[(i)]
        \item Let $C = z I $ be a scalar matrix with $z \in \mathbb{C}$. Then $\mathrm{SRG}_{\theta^\star}(C) \! =\!  z$ with $\theta^\star \! = \! \angle z$, while $\mathrm{SRG}(C) \! = \! \{z,\overline{z}\}$.  
        \item Let $C\in \mathbb{C}^{n\times n}$ be a unitary matrix with $\Lambda(C) = \{ e^{j\phi_1}, \dots, e^{j\phi_n} \}$ satisfying $0 \leq \phi_1 - \phi_n \leq \pi$. Then 
        \begin{equation*}
            \mathrm{SRG}_{\theta^\star}(C) \! =\! \{ e^{j\phi}\!\in\!\mathbb{C}: \phi \!\in\! [\phi_n,\theta^\star \!-\! \delta] \! \cup \! [\theta^\star\!+\!\delta,\phi_1] \}   
        \end{equation*} 
        with $\theta^\star = (\phi_1+\phi_n)/2$ and $ \delta = \min_{i\in\{1,\dots,n\}} |\theta^\star-\phi_i| $. 
        The standard SRG of $C$ is 
        \begin{equation*}
            \mathrm{SRG}(C) \!=\! \{ e^{j\phi} \!\in\! \mathbb{C}\!: \!\phi \!\in\! [-\phi_{\max},-\phi_{\min}]  \cup  [\phi_{\min},\phi_{\max}] \} 
        \end{equation*} 
        with $\phi_{\max} \!=\! \max_{i  \in\! \{1,\dots,n \}} |\phi_i|, \phi_{\min} \!=\! \min_{i \in\! \{1,\dots,n\}} |\phi_i|$. 
         \item Let $C \in \mathbb{C}^{n\times n}$ be a positive definite matrix. Then $\theta^\star = 0$ and $\mathrm{SRG}_{\theta^\star} (C) = \mathrm{SRG}(C)$.  
    \end{enumerate}
\end{example}

\begin{remark} 
    We mention that several approaches can be applied to the computational aspects of the $\theta$-symmetric SRG. In particular, \cite{Huang2019SRG_Normal,Pates2021Scaled} show that the SRG of a matrix can be equivalently characterized via the numerical range of an associated matrix, and \cite{Zhang2025DW_Shell} proposes an SDP-based algorithm for $\theta$-symmetric SRG visualization. 
\end{remark} 

Some properties of $\theta$-symmetric SRG can be derived as follows. 

\begin{lemma}             \label{lem:properties_theta_SRG} 
    Let matrix $C\in \mathbb{C}^{n\times n}$ and $\theta \in \mathbb{R}$  be given, then the following statements are true:  
    \begin{enumerate}[(i)] 
        \item \label{item:theta_SRG_property_1}  Spectrum containment $\Lambda(C) \subset \mathrm{SRG}_\theta(C)$ holds.  
        \item \label{item:theta_SRG_property_2}  Let real $0 \neq \mu \in \mathbb{R}$, then 
        \begin{align*}
            \mathrm{SRG}_\theta(\mu C) & = \mu \mathrm{SRG}_\theta(C) , \\ 
            \mathrm{SRG}_\theta (C + \mu e^{j\theta} I) & = \mathrm{SRG}_\theta(C) + \mu e^{j\theta} . 
        \end{align*} 
        \item \label{item:theta_SRG_property_3}  There holds 
        \begin{equation*} 
            \mathrm{SRG}_{\theta+\pi} (C) = \mathrm{SRG}_\theta (C) . 
        \end{equation*} 
        \item \label{item:theta_SRG_property_4}  If $C$ is nonsingular, then 
        \begin{equation*}
            \mathrm{SRG}_\theta(C^{-1}) = ( \mathrm{SRG}_{-\theta} (C) )^{-1} . 
        \end{equation*} 
    \end{enumerate}
\end{lemma}

\begin{proof} 
    The proof is based on Lemma \ref{lem:property_SRG}. For \eqref{item:theta_SRG_property_1}, note that $\Lambda(e^{-j\theta}C)\subset\mathrm{SRG}(e^{-j\theta}C)$, implying 
    \begin{equation*}
        \Lambda(C) \subset e^{j\theta} \mathrm{SRG}(e^{-j\theta}C) = \mathrm{SRG}_\theta(C).
    \end{equation*} 
    For \eqref{item:theta_SRG_property_2}, there hold 
    $ \mathrm{SRG}_\theta(\mu C) = e^{j\theta} \mathrm{SRG}(\mu e^{-j\theta}C)   = \mu e^{j\theta} \mathrm{SRG}(e^{-j\theta} C) = \mu \mathrm{SRG}_\theta(C)$ and  
    \begin{equation*}
        \mathrm{SRG}_\theta(C\!+\!\mu e^{j\theta} I) \!=\! e^{j\theta} \mathrm{SRG} (e^{-j\theta} C \!+\! \mu I) \! =\! \mathrm{SRG}_\theta(C) \!+\! \mu e^{j\theta} . 
    \end{equation*} 
    For \eqref{item:theta_SRG_property_3},  there holds 
    \begin{align*}
        \mathrm{SRG}_{\theta+\pi} (C)  &  = e^{j(\theta+\pi)}\mathrm{SRG}(e^{-j(\theta+\pi)}C)  \\ &  = - e^{j\theta} \mathrm{SRG}(-e^{-j\theta} C) = \mathrm{SRG}_\theta(C) , 
    \end{align*} 
     where the last equality follows from \eqref{item:theta_SRG_property_2}. \\ 
    To show \eqref{item:theta_SRG_property_4}, we have 
    \begin{align*}
        \mathrm{SRG}_\theta(C^{-1}) & = e^{j\theta} \mathrm{SRG} (e^{-j\theta} C^{-1}) \\ 
        & = (e^{-j\theta} \mathrm{SRG}(e^{j\theta}C))^{-1} = (\mathrm{SRG}_{-\theta} (C) ) ^{-1} . 
    \end{align*} 
    This completes the proof. 
\end{proof}

We can further define the $\theta$-symmetric SRG of a matrix set and establish the subadditive and submultiplicative properties. Given a matrix set $\mathcal{C} \subset \mathbb{C}^{n\times n}$, the $\theta$-symmetric SRG of $\mathcal{C}$ is defined by 
\begin{equation*}
    \mathrm{SRG}_\theta (\mathcal{C}) = \bigcup_{C\in\mathcal{C}} \mathrm{SRG}_\theta(C) . 
\end{equation*} 
Note that a matrix set can be a single matrix. 

The following definition is an extension of the chord property in SRG. 
\begin{definition}[$\theta$-chord property]             \label{defn:theta_chord_property} 
    Let matrix set $\mathcal{C} \! \subset \!  \mathbb{C}^{n\times n}$ and $\theta \in \mathbb{R}$ be given, then $\mathcal{C}$ is said to satisfy the $\theta$-chord property if 
    \begin{equation*}
         \left[ z,\overline{z} e^{2j\theta} \right] \subset \mathrm{SRG}_\theta(\mathcal{C}), \quad \forall z \in \mathrm{SRG}_\theta(\mathcal{C}). 
    \end{equation*} 
\end{definition}

\begin{proposition}             \label{prop:subaddition_theta_SRG}
    Let matrix sets $\mathcal{A},\mathcal{B}\subset\mathbb{C}^{n\times n}$ be given, if there exists $\theta \in \mathbb{R}$ such that $\mathcal{B}$ satisfies the $\theta$-chord property, then 
    \begin{equation}            \label{eq:subaddition_theta_SRG} 
        \mathrm{SRG}_\theta(\mathcal{A}+\mathcal{B}) \subset \mathrm{SRG}_\theta(\mathcal{A}) + \mathrm{SRG}_\theta(\mathcal{B}) .  
    \end{equation} 
\end{proposition}

\begin{proof}
    Suppose $\mathcal{B}$ satisfies the $\theta$-chord property, then for all $z \in \mathrm{SRG}_\theta(\mathcal{B})$, 
    \begin{equation*}
        \left[ z,\overline{z}e^{2j\theta} \right] \subset \mathrm{SRG}_\theta(\mathcal{B}) = e^{j\theta}\mathrm{SRG}\left(e^{-j\theta}\mathcal{B}\right) , 
    \end{equation*} implying 
    \begin{equation*}
        \left[ze^{-j\theta},\overline{z}e^{j\theta}\right] \subset \mathrm{SRG}\left(e^{-j\theta}\mathcal{B}\right). 
    \end{equation*} Note that $z e^{-j\theta} \in \mathrm{SRG}\left(e^{-j\theta}\mathcal{B}\right)$, hence $e^{-j\theta}\mathcal{B}$ satisfies the chord property. By Lemma \ref{lem:SRG_chord_arc_property}, we have 
    \begin{equation}            \label{eq:subaddition_proof_inclusion}
        \mathrm{SRG}(e^{-j\theta}\mathcal{A}+e^{-j\theta}\mathcal{B}) \subset \mathrm{SRG}(e^{-j\theta}\mathcal{A}) + \mathrm{SRG}(e^{-j\theta}\mathcal{B}). 
    \end{equation} Multiplying both sides of \eqref{eq:subaddition_proof_inclusion} by $e^{j\theta}$ yields \eqref{eq:subaddition_theta_SRG} holds. This completes the proof.  
\end{proof}

Define arcs symmetric with respect to the line $\{ \mu e^{j\theta}:\mu\in\mathbb{R} \}$ as 
\begin{equation*} 
    \mathrm{Arc}_\theta^+ (z,\overline{z}e^{2j\theta}) = \{ |z| e^{j(\mu\angle z+\theta(1-\mu))}:  \mu \in[-1,1]  \} , 
\end{equation*} 
and 
\begin{equation*}
    \mathrm{Arc}_\theta^- (z,\overline{z}e^{2j\theta}) = - \mathrm{Arc}_\theta^+ (-z,-\overline{z}e^{2j\theta}) . 
\end{equation*} 
Then the following definition is an extension of the arc property in SRG. 

\begin{definition}[$\theta$-arc property]  
\label{defn:theta_arc_property} 
Let matrix set $\mathcal{C} \subset \mathbb{C}^{n\times n}$ and $\theta \in \mathbb{R}$ be given, then $\mathcal{C}$ is said to satisfy the $\theta$-arc property if either of the following holds 
\begin{enumerate}[(i)]
    \item $ \mathrm{Arc_\theta^+}(z,\overline{z}e^{2j\theta}) \subset \mathrm{SRG}_\theta(\mathcal{C}), \quad \forall z \in \mathrm{SRG}_\theta(\mathcal{C}) $. 
    \item $ \mathrm{Arc_\theta^-}(z,\overline{z}e^{2j\theta}) \subset \mathrm{SRG}_\theta(\mathcal{C}), \quad \forall z \in \mathrm{SRG}_\theta(\mathcal{C}) $. 
\end{enumerate} 
\end{definition}

\begin{proposition}               \label{prop:submultiplication_theta_SRG}
    Let matrix sets $\mathcal{A},\mathcal{B} \subset \mathbb{C}^{n\times n}$ be given, if there exists $\beta\in \mathbb{R}$ such that $\mathcal{B}$ satisfies the $\beta$-arc property, then 
    \begin{equation}                   \label{eq:submultiplication_theta_SRG}
        \mathrm{SRG}_{\alpha+\beta}(\mathcal{A}\mathcal{B}) \subset \mathrm{SRG}_{\alpha} (\mathcal{A}) \mathrm{SRG}_{\beta} (\mathcal{B}) 
    \end{equation} 
    for any $\alpha \in \mathbb{R}$. 
\end{proposition}

\begin{proof}
    Suppose $\mathcal{B}$ satisfies $\beta$-arc property. Without loss of generality, we assume $\mathrm{Arc_{\beta}^+}(z,\overline{z}e^{2j\beta}) \subset \mathrm{SRG}_{\beta}(\mathcal{B})$ for all $z \in \mathrm{SRG}_{\beta}(\mathcal{B})$ (the case where $\mathrm{Arc_{\beta}^-}(z,\overline{z}e^{2j\beta}) \subset \mathrm{SRG}_{\beta}(\mathcal{B})$ follows by a similar argument). Then 
    \begin{align*} 
        e^{-j\beta}\mathrm{Arc}_{\beta}^+(z, \overline{z}e^{2j\beta}) & = \mathrm{Arc}_{\beta}^+(z e^{-j\beta} , \overline{z}e^{j\beta}) \\ & \subset \mathrm{SRG}(e^{-j\beta}\mathcal{B}) , 
    \end{align*} implying $e^{-j\beta}\mathcal{B}$ satisfies the arc property since $ze^{-j\beta}  \in \mathrm{SRG}(e^{-j\beta}\mathcal{B})$. Hence it follows from Lemma \ref{lem:SRG_chord_arc_property} that for any $\alpha \in \mathbb{R}$, there holds 
    \begin{equation}            \label{eq:submultiplication_proof_inclusion}
        \mathrm{SRG}(e^{-j(\alpha+\beta)}\mathcal{A}\mathcal{B}) \subset \mathrm{SRG}(e^{-j\alpha}\mathcal{A}) \mathrm{SRG} (e^{-j\beta}\mathcal{B}) . 
    \end{equation} 
    Multiplying both sides of \eqref{eq:submultiplication_proof_inclusion} by $e^{j(\alpha+\beta)}$ yields \eqref{eq:submultiplication_theta_SRG} holds. This completes the proof. 
\end{proof}

\subsection{$\theta$-Symmetric SRG of Cascaded Product versus Product of Corresponding $\theta$-Symmetric SRGs}

We are interested in whether $-1$ is an eigenvalue of the cascaded product of matrices $A_1A_2\cdots A_N$, which is important in the stability analysis of cyclic feedback systems. The following is the main result in this section, providing a graphical criterion for the singularity of $I+A_1A_2\cdots A_N$ via $\theta$-symmetric SRG. 

\begin{theorem}                     \label{thm:matrix_cyclic_interconnection}
    Let $A_1,A_2,\dots,A_N \in \mathbb{C}^{n\times n}$ be given, then $I+A_1A_2\cdots A_N$ is nonsingular if there exists $\alpha_i \in \mathbb{R}, i = 1,2,\dots, N$, such that 
    \begin{equation}                 \label{eq:cyclic_matrix_condition} 
         -1 \notin  \prod_{i=1}^N \mathrm{SRG}_{\alpha_i}(\mathcal{A}_i),  
    \end{equation} 
    where each $\mathcal{A}_i$ is an arbitrary matrix set containing $A_i$, and at least $N-1$ of the $\mathcal{A}_i$ satisfy $\alpha_i$-arc property. 
\end{theorem}

\begin{proof}
    Suppose, without loss of generality, that there exists an index $k \in \{ 1,2,\dots, N \}$ such that $\mathcal{A}_k$ does not satisfy $\alpha_k$-arc property, while $\mathcal{A}_i$ satisfies $\alpha_i$-arc property for all $i \neq k$. Since a cyclic permutation does not affect the singularity of $I+A_1A_2\cdots A_N$, it follows that $I+A_1A_2\cdots A_N$ is nonsingular if and only if $I+A_k A_{k+1}\cdots A_NA_1\cdots A_{k-1}$ is nonsingular. Then 
    \begin{align*}
        & \Lambda(A_k A_{k+1}\cdots A_NA_1\cdots A_{k-1}) \\ & \qquad \qquad    \subset \Lambda( \mathcal{A}_k \mathcal{A}_{k+1} \cdots \mathcal{A}_N \mathcal{A}_1 \cdots \mathcal{A}_{k-1} ) \\ & \qquad \qquad  \subset \mathrm{SRG}_{\sum_{i=1}^N \alpha_i} (\mathcal{A}_k \mathcal{A}_{k+1} \cdots \mathcal{A}_N \mathcal{A}_1 \cdots \mathcal{A}_{k-1}) \\ 
        & \qquad \qquad  \subset  \prod_{i=1}^N \mathrm{SRG}_{\alpha_i}(\mathcal{A}_i) . 
    \end{align*} 
    In view of \eqref{eq:cyclic_matrix_condition}, we have $-1$ is not an eigenvalue of $A_k A_{k+1}\cdots A_NA_1\cdots A_{k-1}$. Hence $I+A_1A_2\cdots A_N$ is nonsingular. This completes the proof. 
\end{proof}

Fixing $\alpha_i = 0$, we obtain the following corollary, which coincides with the SRG result in \cite{Chaffey2023Graphical}. The proof follows directly from Theorem \ref{thm:matrix_cyclic_interconnection} and is omitted. 

\begin{corollary}                       \label{cor:reduce_to_cyclic_SRG}
    Let $A_1,A_2,\dots,A_N \in \mathbb{C}^{n\times n}$ be given, then $I+A_1A_2\cdots A_N$ is nonsingular if 
    \begin{equation*}
        -1 \notin  \prod_{i=1}^N \mathrm{SRG}(\mathcal{A}_i) , 
    \end{equation*} 
    where each $\mathcal{A}_i$ is an arbitrary matrix set containing $A_i$, and at least $N-1$ of the $\mathcal{A}_i$ satisfy the arc property. 
\end{corollary}

By Theorem \ref{thm:matrix_cyclic_interconnection}, a natural next step is how to choose the optimal $\alpha_i$ to maximize the distance between $-1$ and the product of $\theta$-symmetric SRGs. To be precise, let $a = \{ \alpha_1, \alpha_2,\dots, \alpha_N \} $ and consider the optimization problem 
\begin{equation}                                \label{eq:find_optimal_alpha}
    a^\star = \argmax_{a \in \mathbb{R}^N} \min_{\substack{s_i \in \mathrm{SRG}_{\alpha_i}(\mathcal{A}_i), \\ i \in \{ 1,2,\dots,N \} } } \left|1+\prod_{i=1}^N s_i \right| . 
\end{equation} 
This problem is analogous in spirit to the block-diagonal $D$-scaling used in structured singular value ($\mu$) analysis \cite{Zhou1996Robust}, where one seeks a block-diagonal matrix $D$ that minimizes $\overline{\sigma}(DMD^{-1})$, thereby tightening an upper bound of $\mu_\Delta(M)$. Another related study is \cite{Chen2025Cyclic}, which similarly introduces a phase-based scaling problem. Note that choosing $\alpha_i$ to minimize the phase spread of each $\mathrm{SRG}_{\alpha_i}(\mathcal{A}_i)$ separately does not, in general, solve the problem \eqref{eq:find_optimal_alpha}. In fact, \eqref{eq:find_optimal_alpha} accounts for the interactions among different $\mathrm{SRG}_{\alpha_i}(\mathcal{A}_i)$ and is a mixed gain and phase optimization problem, which is the subject of ongoing work.

When restricting to the two components case, we focus on the singularity of matrix $I+AB$. The following result follows from Theorem \ref{thm:matrix_cyclic_interconnection} and \cite[Theorem IV.2]{Zhang2025DW_Shell}, and therefore its proof is omitted. 

\begin{corollary}                       \label{cor:two_components_matrix_condition}
    Let $A,B \in \mathbb{C}^{n\times n}$ be given with $A$ nonsingular, then $I+AB$ is nonsingular if any of the following conditions is satisfied: 
    \begin{enumerate}[(i)]
        \item \label{item:two_component_matrix_condition_1} There exists $\theta \in \mathbb{R}$ such that 
        \begin{equation*}
            (\mathrm{SRG}_\theta(A))^{-1} \cap \mathrm{SRG}_\theta(-B) = \emptyset. 
        \end{equation*} 
        \item \label{item:two_component_matrix_condition_2} There exists $\theta \in \mathbb{R}$ such that 
        \begin{equation*}                    
            0 \notin \mathrm{SRG}_\theta(A^{-1}) + \mathrm{SRG}_\theta(\mathcal{B}_\mathrm{chord}), 
        \end{equation*}  
        where $\mathcal{B}_\mathrm{chord}$ is an arbitrary matrix set containing $B$ and satisfying $\theta$-chord property. 
        \item \label{item:two_component_matrix_condition_3} There exist $\alpha,\beta \in \mathbb{R}$ such that 
        \begin{equation*}                     
            -1 \notin \mathrm{SRG}_\alpha (A) \mathrm{SRG}_\beta(\mathcal{B}_\mathrm{arc}) ,
        \end{equation*} 
        where $\mathcal{B}_\mathrm{arc}$ is an arbitrary matrix set containing $B$ and satisfying $\beta$-arc property.  
    \end{enumerate}
\end{corollary}

Some connections and differences of the method in Theorem \ref{thm:matrix_cyclic_interconnection} and the graph separation method can be inferred from Corollary \ref{cor:two_components_matrix_condition}. Note that conditions \eqref{item:two_component_matrix_condition_1} and \eqref{item:two_component_matrix_condition_2} in Corollary \ref{cor:two_components_matrix_condition} are standard graph separation methods for the two-component case. Moreover, \eqref{item:two_component_matrix_condition_1} enjoys lower conservatism than \eqref{item:two_component_matrix_condition_2}, since the latter over-approximates $B$ by using a superset. In contrast, condition \eqref{item:two_component_matrix_condition_3} adopts a fundamentally different approach from graph separation. It operates directly on the matrix product, rather than dealing with inverse matrices. This is instrumental in handling the cyclic case, where the standard graph separation method is not applicable.

\subsection{Over-approximations of Matrices via $\theta$-Symmetric SRG}          \label{subsection:over_approximation}

In the previous analysis, the over-approximation of a matrix via its $\theta$-symmetric SRG, namely a superset containing the matrix itself, is important for establishing the main result. Here we analyze some typical over-approximations with intuitive graphical interpretation, which helps to draw explicit connections to the results in existing literature.

A single matrix generally does not satisfy the $\theta$-arc property. Hence a natural approach is to over-approximate it by matrix sets possessing the $\theta$-arc property. Motivated by this, we introduce the notion of arc hull of the $\theta$-symmetric SRG. For a matrix $C \in \mathbb{C}^{n \times n}$ and $\theta \in \mathbb{R}$, denote  
\begin{align*}
        \mathrm{Hull}_{\mathrm{arc}} (\mathrm{SRG}_\theta(C))   \!  = \! \{ \mathrm{Arc}_\theta^+ \mspace{-2mu} (z,\mspace{-1mu} \overline{z}e^{2j\theta}) \mspace{-1mu} \! \subset \!  \mathbb{C}\! :\! z \! \in \!  \mathrm{SRG}_\theta(C)  \} . 
\end{align*}
Accordingly, we can graphically define the following matrix set via the $\theta$-symmetric SRG: 
\begin{equation}                 \label{eq:arc_approximation_set}  
    \mathcal{M}_{\theta\!-\!\mathrm{arc}}(C) \mspace{-1mu} \! = \! \mspace{-1mu} \{ M \! \mspace{-1mu} \in \! \mathbb{C}^{n\!\times\! n} \mspace{-1mu} \! :\!   \mathrm{SRG}_\theta(M) \mspace{-1mu} \! \subset \! \mspace{-1mu} \mathrm{Hull}_{\mathrm{arc}} (\mathrm{SRG}_\theta(C))  \} .  
\end{equation}

Clearly,  $\mathcal{M}_{\theta-\mathrm{arc}}(C)$ is an over-approximation of the matrix $C$. The following result shows that $\mathcal{M}_{\theta-\mathrm{arc}}(C)$ satisfies the $\theta$-arc property.

\begin{lemma}                   \label{lem:theta_arc_approximation}
    Let matrix $C \in \mathbb{C}^{n\times n}$ and $\theta \in \mathbb{R}$ be given. Then 
    $\mathcal{M}_{\theta-\mathrm{arc}}(C)$ defined in \eqref{eq:arc_approximation_set} satisfies the $\theta$-arc property.  
\end{lemma} 

\begin{proof}
    Note that, by definition, 
    \begin{equation}                  \label{eq:approximation_theta_arc}  
        \mathrm{SRG}_\theta ( \mathcal{M}_{\theta-\mathrm{arc}}(C) )  \subset  \mathrm{Hull}_{\mathrm{arc}} (\mathrm{SRG}_\theta(C))  .   
    \end{equation}  
    It remains to show the reverse inclusion. Consider a subset of $\mathcal{M}_{\theta-\mathrm{arc}} (C)$ defined by 
    \begin{equation*}
        \mathcal{M} \! = \! \{ z I \! \in \! \mathbb{C}^{n\times n}\!: z \! \in \!  \mathrm{Hull}_{\mathrm{arc}} (\mathrm{SRG}_\theta(C))  \} . 
    \end{equation*}
    Then there holds  $\mathrm{SRG}_\theta(\mathcal{M}) \! = \! \mathrm{Hull}_{\mathrm{arc}} (\mathrm{SRG}_\theta(C))  $, implying $ \mathrm{SRG}_\theta(\mathcal{M}_{\theta-\mathrm{arc}} (C) ) = \mathrm{Hull}_{\mathrm{arc}} (\mathrm{SRG}_\theta(C)) $ in view of \eqref{eq:approximation_theta_arc}. Hence $\mathcal{M}_{\theta-\mathrm{arc}} (C) $ satisfies the $\theta$-arc property. 
\end{proof}

\begin{remark}              \label{rem:similar_holds_for_left_arc} 
    Analogous arguments apply when considering the arc hull of $\theta$-symmetric SRG defined by $\mathrm{Arc}_\theta^-(z,\overline{z}e^{2j\theta})$. 
\end{remark}

The following result provides a criterion for the singularity of $I+A_1A_2\cdots A_N$ in terms of over-approximations of the matrices. The proof follows from Lemma \ref{lem:theta_arc_approximation} and Theorem \ref{thm:matrix_cyclic_interconnection}, hence it is omitted for brevity. 

\begin{corollary}               \label{cor:arc_hull_result}
     Let $A_1,A_2,\dots,A_N \in \mathbb{C}^{n\times n}$ be given, then $I+A_1A_2\cdots A_N$ is nonsingular if there exists $\alpha_i \in \mathbb{R}, i = 1,2,\dots, N$, and an index $k \in \{1,\dots,N\}$ such that 
    \begin{equation}                 \label{eq:arc_hull_result}  
        -1 \notin \mathrm{SRG}_{\alpha_k} (A_k) \prod_{i=1,i\neq k}^N \mathrm{Hull}_{\mathrm{arc}}(\mathrm{SRG}_{\alpha_i}(A_i)) . 
    \end{equation}   
\end{corollary}

To facilitate connections with the results in existing literature, we first review for preparation the notion of $\theta$-segmental phase \cite{Chen2025Cyclic}. 
For a matrix $C \in \mathbb{C}^{n\times n}$ and $\theta \in \mathbb{R}$, the $\theta$-segmental phase $\Psi_\theta(C)$ is defined as 
\begin{equation}                    \label{eq:theta_segmental_phase}
    \Psi_\theta(C) = \left[\underline{\psi}_\theta(C), \overline{\psi}_\theta(C) \right], 
\end{equation} 
where 
\begin{align*}
    \overline{\psi}_\theta(C) & = \theta + \Gamma_\theta(C) , \\  
    \underline{\psi}_\theta(C) & = \theta - \Gamma_\theta(C) ,  
\end{align*} 
with $\Gamma_\theta(C)$ given in \eqref{eq:phase_spread}. When $\theta = 0$, then 
\begin{equation*}
    \Psi_0(C) = [ - \Gamma_0(C), \Gamma_0(C) ] , 
\end{equation*} 
where $\Gamma_0(C)$ is called the singular angle of $C$ \cite{Wielandt1967Topics}. When $\theta^\star$ is determined by \eqref{eq:optimal_theta}, the corresponding $\Psi_{\theta^\star}(C)$ defines the segmental phase of $C$, which attains the minimal phase spread. 

Now, define a subset of $\mathbb{C}$ as 
\begin{equation*}
    \mathscr{S}_\theta(C) \! \mspace{-1mu} = \mspace{-1mu} \! \big\{ z \! \in \! \mathbb{C}\! : \! |z| \!\in \! [\underline{\sigma}(C), \overline{\sigma}(C)],    \angle z \! \in \! \big[\underline{\psi}_\theta   (C), \overline{\psi}_\theta  (C)\big] \big\} , 
\end{equation*} 
which corresponds to an annular sector in the complex plane. Then there holds 
\begin{equation}                \label{eq:approximation_connection}
    \mathrm{SRG}_\theta(C) \subset \mathrm{Hull}_{\mathrm{arc}}(\mathrm{SRG}_\theta(C)) \subset \mathscr{S}_\theta(C). 
\end{equation} 
The following example provides an intuitive illustration of this connection. 

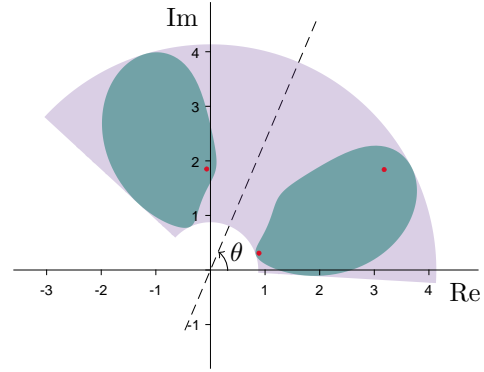
\begin{figure}[htb]
    \centering
    \begin{overpic}[width=0.7\linewidth]{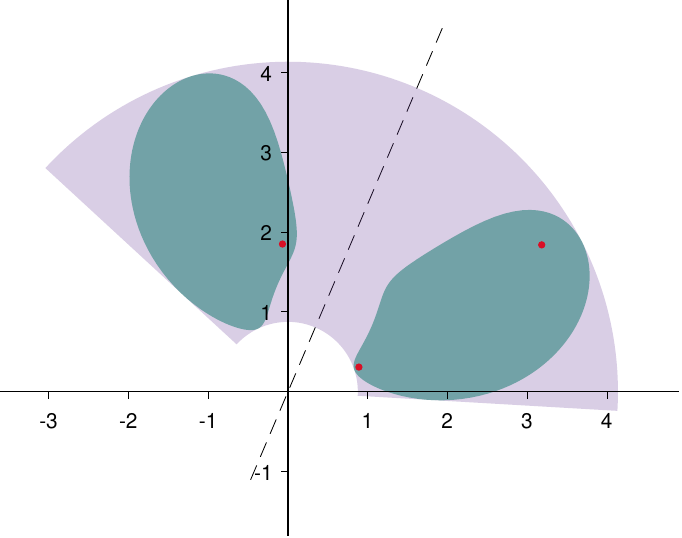}
        \put(82,41){$\theta$}   
        \put(165,26){$\mathrm{Re}$} 
        \put(58,130){$\mathrm{Im}$} 
        \begin{tikzpicture}[overlay] 
            \draw[->] (2.85,1.315) arc(0:52:0.32);  
        \end{tikzpicture} 
    \end{overpic}
    \caption{An illustration of $\mathrm{SRG}_\theta(C)$ (cyan-green region) and its over-approximation $\mathscr{S}_\theta(C)$ (purple region) with $\theta = 67^\circ$, where the red points are the eigenvalues of $C$.}
    \label{fig:SRG_theta_over_approximation} 
\end{figure}

\begin{example}
    Consider a matrix 
    \begin{equation*}
        C =  \begin{bmatrix}
            2j & 0 & 1 \\ -1+j & 2+j& j \\ j & 1-j & 2+j 
        \end{bmatrix} . 
    \end{equation*} 
    Let $\theta=67^\circ$, then an illustration of the connection \eqref{eq:approximation_connection} is shown in Fig. \ref{fig:SRG_theta_over_approximation}. 
\end{example}

Hence, by Corollary \ref{cor:arc_hull_result}, we have the following result. 

\begin{corollary}           \label{cor:over_approximation_nonsingularity}
    Let $A_1,A_2,\dots,A_N \in \mathbb{C}^{n\times n}$ be given, then $I+A_1A_2\cdots A_N$ is nonsingular if there exists $\alpha_i \in \mathbb{R}, i = 1,2,\dots, N$, and an index $k \in \{1,\dots,N\}$ such that 
    \begin{equation}                    \label{eq:annular_sector_result}      
        -1 \notin \mathrm{SRG}_{\alpha_k}(A_k) \mathscr{S} 
    \end{equation} 
    where $\mathscr{S}$ is an annular sector defined by 
    \begin{multline*}
        \mathscr{S} \! =\! \left\{ z \! \in \! \mathbb{C}:  |z| \! \in \!  \left[\prod_{i=1,i\neq k}^N \underline{\sigma}(A_i)  , \prod_{i = 1,i \neq k}^N \overline{\sigma}(A_i)\right] \right., \\ \ \  \angle z \! \in \! \left. \left[ \sum_{i=1,i\neq k}^N \underline{\psi}_{\alpha_i}(A_i), \sum_{i=1,i\neq k}^N \overline{\psi}_{\alpha_i}(A_i) \right] \right\} . 
    \end{multline*}  
\end{corollary}

\begin{proof}
Note that multiplication of annular sectors remains an annular sector. In view of \eqref{eq:approximation_connection}, 
\begin{equation*} 
    \prod_{i=1,i\neq k}^N \mathrm{Hull}_{\mathrm{arc}}(\mathrm{SRG}_{\alpha_i}(A_i) \subset 
     \prod_{i=1,i\neq k}^N \mathscr{S}_{\alpha_i}(A_i) = \mathscr{S}.  
\end{equation*} 
Hence, \eqref{eq:annular_sector_result} implies \eqref{eq:arc_hull_result} holds, yielding $I+A_1A_2\cdots A_N$ is nonsingular by Corollary \ref{cor:arc_hull_result}. This completes the proof. 
\end{proof}

When gain and phase are considered separately, Corollary \ref{cor:over_approximation_nonsingularity} degenerates into the following two corollaries. Notably, Corollary \ref{cor:segmental_phase_cyclic_matrix} is consistent with the result in \cite{Chen2025Cyclic}. The proofs are omitted for brevity. 

\begin{corollary}               \label{cor:gain_cyclic_matrix} 
    Let $A_1,A_2,\dots,A_N \in \mathbb{C}^{n\times n}$ be given, then $I+A_1A_2\cdots A_N$ is nonsingular if 
    \begin{equation*}
        \prod_{i = 1}^N \overline{\sigma}(A_i) < 1. 
    \end{equation*}
\end{corollary}

\begin{corollary}               \label{cor:segmental_phase_cyclic_matrix} 
    Let $A_1,A_2,\dots,A_N \in \mathbb{C}^{n\times n}$ be given, then $I+A_1A_2\cdots A_N$ is nonsingular if there exists $\alpha_i \in \mathbb{R}, i = 1,2,\dots, N$, such that 
    \begin{equation*}
        \sum_{i=1}^N \Psi_{\alpha_i}(A_i) \subset (-\pi,\pi). 
    \end{equation*} 
\end{corollary}

\section{Stability Analysis of Cyclic Cascaded Systems via $\theta$-symmetric SRG} 
\label{section:stability_systems}

For a system $G \in \mathcal{RH}_\infty^{m\times m}$, by plotting the $\theta$-symmetric SRG of $G(j\omega)$ at each frequency, we obtain an over-approximation of the eigenloci. This serves as a MIMO extension of the Nyquist plot and is especially well suited for stability analysis of cyclic feedback systems. An illustrative example of the frequency-wise $\theta$-symmetric SRG of a system is provided below.  

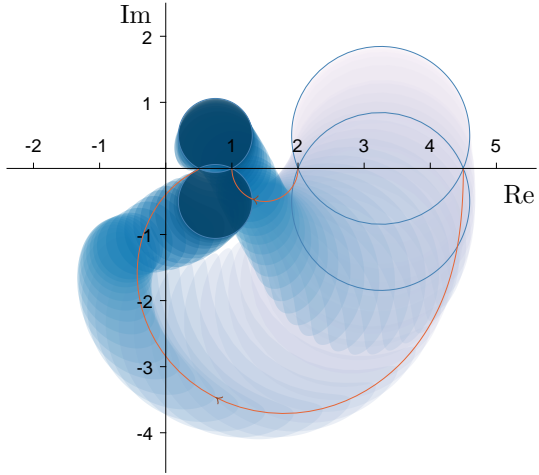
\begin{figure}[htb]
    \centering
    \begin{overpic}[width=0.8\linewidth]{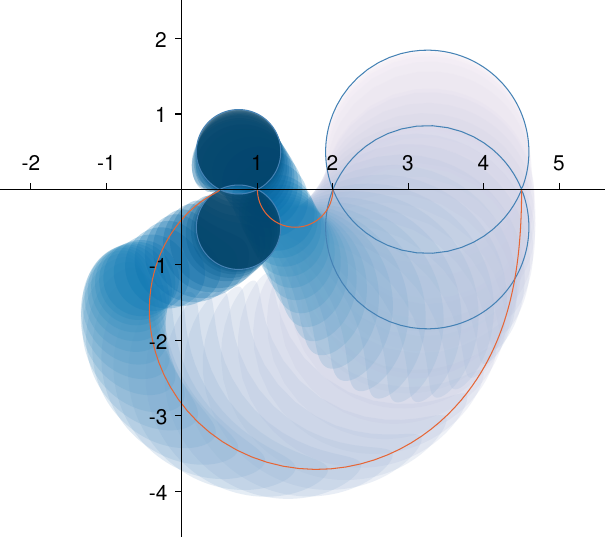}
        \put(188,103){$\mathrm{Re}$} 
        \put(43,171){$\mathrm{Im}$} 
        \begin{tikzpicture}[overlay] 
            \draw[->, color={rgb:red,0.88;green,0.42;blue,0.21}] (2.8,1.023) arc(-100:-125:0.01);  
            \draw[->, color={rgb:red,0.88;green,0.42;blue,0.21}] (3.285,3.668) arc(-70:-100:0.01);  
        \end{tikzpicture} 
    \end{overpic}
    \caption{An illustration of the frequency-wise $\theta$-symmetric SRG of $G(j\omega)$ in \eqref{eq:system_example}, where the red curves denote the eigenloci of $G(j\omega)$. }
    \label{fig:system_frequency_wise_theta_SRG} 
\end{figure}

\begin{example}                     \label{examp:system_theta_SRG_example}
    Consider a system 
    \begin{equation}                \label{eq:system_example} 
        G(s) = \begin{bmatrix}
            \frac{s+2}{s+1} & 0 \\ 1 & \frac{s^2+3s+27}{2s^2+4s+6} 
        \end{bmatrix} . 
    \end{equation} 
    Choose $\theta(\omega) = \angle ( \frac{j\omega+2}{j\omega+1}+\frac{3j\omega+27-\omega^2}{4j\omega+6-2\omega^2} )$. Then the frequency-wise $\theta$-symmetric SRG of $G(j\omega)$ is shown in Fig. \ref{fig:system_frequency_wise_theta_SRG}. From the plot, $G(s)$ exhibits an overall phase-lag behavior, which the standard SRG method fails to capture. 
\end{example}

Now consider the feedback interconnection of a cascade of systems shown in Fig. \ref{fig:cyclic_feedback_system}. Denote 
$e = \begin{bmatrix}
    e_1^T & e_2^T & \cdots & e_N^T
\end{bmatrix}^T$ and 
$u = \begin{bmatrix}
    u_1^T & u_2^T & \cdots& u_N^T
\end{bmatrix}^T$. Then the cyclic feedback system is said to be stable if the transfer matrix from $e$ to $u$ 
\begin{equation*}
    \begin{bmatrix}
        I      & -P_2      &  0  & \cdots &   0    \\
        0      & I     & -P_3      & \ddots & \vdots      \\
        \vdots     & 0  & I      & \ddots & 0 \\
        0 & \vdots & \ddots & \ddots & -P_N    \\
        P_1      & 0 & \cdots      & 0 & I
    \end{bmatrix}^{-1} 
\end{equation*} 
belongs to $\mathcal{RH}_\infty^{Nm\times Nm}$. When $P_i\in \mathcal{RH}_\infty^{m\times m}$, the cyclic feedback system is stable if and only if \cite{Zhou1996Robust,Chen2025Cyclic} 
\begin{equation}                \label{eq:to_show_cyclic_stability_condition}
    (I+P_1P_2\cdots P_N)^{-1} \in \mathcal{RH}_\infty^{m \times m}. 
\end{equation}

Now we are ready to show our main result, which provides a graphical stability criterion for the feedback interconnection of cascaded systems via $\theta$-symmetric SRG. 

\begin{theorem}                         \label{thm:cyclic_system_theta_SRG}
    The feedback interconnection of a cascade of systems $P_1,P_2,\dots,P_N\in \mathcal{RH}_\infty^{m \times m}$ in Fig. \ref{fig:cyclic_feedback_system} is stable if there exist functions $\alpha_i(\omega) \in \mathbb{R}, i = 1, 2, \dots, N$, such that for all $\omega \in [0,\infty]$ and all $\tau \in [0,1]$, 
    \begin{equation}            \label{eq:cyclic_system_stability_condition}
        -1 \notin \tau \prod_{i = 1}^N \mathrm{SRG}_{\alpha_i(\omega)}(\mathcal{P}_i(j\omega)), 
    \end{equation} 
    where each $\mathcal{P}_i(j\omega)$ is an arbitrary matrix set containing $P_i(j\omega)$, and at least $N-1$ of the $\mathcal{P}_i(j\omega)$ satisfy $\alpha_i(\omega)$-arc property. 
\end{theorem}

\begin{proof}
    Since $P_i \in \mathcal{RH}_\infty^{m \times m}$, it suffices to show \eqref{eq:to_show_cyclic_stability_condition} holds. It follows from the generalized Nyquist criterion \cite{Desoer1980Generalized,Skogestad2005MultivariableFeedbackControl,Griggs2012InterconnectionsMixed} that \eqref{eq:to_show_cyclic_stability_condition} holds if 
    \begin{equation}           \label{eq:sufficient_nyquist_condition}
        I + \tau P_1(j\omega)P_2(j\omega)\cdots P_N(j\omega) \ \text{ is nonsingular,   } 
    \end{equation} 
    for all $\omega \in [0,\infty]$ and all $\tau\in[0,1]$. \\ 
    By Theorem \ref{thm:matrix_cyclic_interconnection}, \eqref{eq:cyclic_system_stability_condition} implies that for any $\omega \in [0,\infty]$, 
    \begin{equation*}
        -1 \notin \tau \Lambda(P_1(j\omega)P_2(j\omega)\cdots P_N(j\omega)) , \quad \forall \tau \in [0,1]. 
    \end{equation*} 
    Consequently, this means that $-1$ is not an eigenvalue of $\tau  P_1 (j\omega) P_2(j\omega) \cdots P_N (j\omega) $ and \eqref{eq:sufficient_nyquist_condition} holds. Hence the cyclic feedback interconnection is stable and the proof is completed.  
\end{proof} 

\begin{remark}              \label{rem:alpha_dependent_on_tau_omega} 
    As noted in \cite[Remark V.1]{Zhang2025DW_Shell}, the parameter $\alpha_i$ in Theorem \ref{thm:cyclic_system_theta_SRG} can be selected as a function jointly dependent on both $\tau$ and $\omega$, thereby enabling a more flexible and less conservative condition.   
\end{remark}

By fixing $\alpha_i(\omega) = 0$, Theorem \ref{thm:cyclic_system_theta_SRG} reduces to the following standard SRG result. The proof is omitted for brevity. 

\begin{corollary}
    The feedback interconnection of a cascade of systems $P_1,P_2,\dots,P_N \in \mathcal{RH}_\infty^{m\times m}$ in Fig. \ref{fig:cyclic_feedback_system} is stable if for all $\omega \in [0,\infty]$ and all $\tau \in [0,1]$, 
    \begin{equation*}
        -1 \notin \tau \prod_{i=1}^N\mathrm{SRG}(P_i(j\omega)), 
    \end{equation*} 
    where each $\mathcal{P}_i(j\omega)$ is an arbitrary matrix set containing $P_i(j\omega)$, and at least $N-1$ of the $\mathcal{P}_i(j\omega)$ satisfy arc property. 
\end{corollary}

Based on an over-approximation of the $\theta$-symmetric SRG in Section \ref{subsection:over_approximation}, we obtain the following result, which can be reduced to \cite[Theorem 4]{Chen2025Cyclic}. Since it follows directly from Corollary \ref{cor:gain_cyclic_matrix} and Corollary \ref{cor:segmental_phase_cyclic_matrix}, the proof is omitted. 

\begin{proposition}               \label{prop:cyclic_system_mixed_gain_phase}
        The feedback interconnection of a cascade of systems $P_1,P_2,\dots,P_N \in \mathcal{RH}_\infty^{m\times m}$ in Fig. \ref{fig:cyclic_feedback_system} is stable if for each $\omega \in [0,\infty]$, one of the following conditions holds: 
        \begin{enumerate}[(i)]
            \item $\displaystyle  \prod_{i=1}^N \overline{\sigma}(P_i(j\omega)) < 1$; 
            \item there exist $\alpha_i(\omega) \in \mathbb{R}, i = 1,2,\dots,N$, such that 
            \begin{equation*}
                \sum_{i=1}^N \Psi_{\alpha_i(\omega)} (P_i(j\omega)) \subset (-\pi,\pi) . 
            \end{equation*}
        \end{enumerate}
\end{proposition}

\section{Numerical Examples} 
\label{section:examples} 

Here we give two examples, one to illustrate the gap between $\theta$-symmetric SRG and the standard SRG; the other to show the effectiveness of $\theta$-symmetric SRG when dealing with cyclic interconnections of MIMO LTI systems. 

\textbf{Example 6. } In this example, we will show the gap between $\theta$-symmetric SRG and the standard SRG. Consider two SISO transfer functions 
\begin{equation*}
    p_1(s) = \frac{s^2+4s+36}{s^2+3s+5}, \quad p_2(s) = \frac{1}{p_1(s) } = \frac{s^2+3s+5}{s^2+4s+36} . 
\end{equation*}  
Clearly, $\left(1+p_1(s)p_2(s)\right)^{-1} \in \mathcal{RH}_\infty$, i.e, the feedback system of $p_1(s)$ and $p_2(s)$ is stable. 

Note that for $\omega_0 = 3$, $p_1(j\omega_0) = -3j$. By \eqref{eq:SRG_matrix}, the SRG of $p_1(j\omega_0)$ and $p_2(j\omega_0)$ are 
\begin{equation*}
    \mathrm{SRG}(p_1(j\omega_0)) \! = \!  \{  -3j, 3j \}, \ \  \mathrm{SRG}(p_2(j\omega_0)) \! = \!  \{ \frac{1}{3}j,-\frac{1}{3}j  \} , 
\end{equation*}
implying 
\begin{equation*}
    -1 \in \mathrm{SRG}(p_1(j\omega_0)) \mathrm{SRG}(p_2(j\omega_0)) = \{ 1, -1 \}. 
\end{equation*} 
Hence, the SRG based methods are not applicable. 

Now we show the effectiveness of Theorem \ref{thm:cyclic_system_theta_SRG}. For every frequency $\omega\in[0,\infty]$, denote $p_1(j\omega) = r(\omega)e^{j\phi(\omega)}$, then $p_2(j\omega) \! = \!  \frac{1}{r(\omega)} e^{-j\phi(\omega)}$. Choose 
$ \alpha_1(\omega) \! = \!  \phi(\omega) $ and $ \alpha_2(\omega) \! = \!  -\phi(\omega) $. 
It follows from \eqref{eq:definition_PS_SRG_matrix} that 
\begin{equation*}
    \mathrm{SRG}_{\alpha_1(\omega)}(p_1(j\omega)) \! = \!  p_1(j\omega),  \ \ \mathrm{SRG}_{\alpha_2(\omega)}(p_2(j\omega)) \! = \! p_2(j\omega)  .  
\end{equation*}  Hence, 
\begin{align*}
    -1 \notin \tau \mathrm{SRG}_{\alpha_1(\omega)}(p_1(j\omega)) \mathrm{SRG}_{\alpha_2(\omega)}(p_2(j\omega)) = \tau , 
\end{align*} 
for all $\tau \in [0,1]$. Note that $p_2(j\omega)$ itself satisfies the $\alpha_2(\omega)$-arc property. Hence, by Theorem \ref{thm:cyclic_system_theta_SRG}, the feedback system of $p_1(s)$ and $p_2(s)$ is stable. 

From the above example, one can see the limitation of the standard SRG arises from the inclusion of the conjugate terms, which increases conservatism and fails to handle some SISO cases. In contrast, the $\theta$-symmetric SRG resolves this by refining the phase information of the system, resulting in a more effective and less conservative framework. 

\textbf{Example 7. } Here, we show the effectiveness of $\theta$-symmetric SRG to deal with cyclic interconnection of MIMO LTI systems. Consider three systems 
$ P_1(s) = \begin{bmatrix}
        1  & 0  \\  \frac{s+4}{2s+3} & \frac{s+2}{s+1} 
    \end{bmatrix},   P_2(s) = \begin{bmatrix}
        \frac{s+1}{2s+5} & 0 \\ \frac{1}{2s+2} & 1 
    \end{bmatrix}, $ and 
    $ P_3(s) = \begin{bmatrix}
        1 & - \frac{1}{2s+2}  \\  0 & \frac{s+2}{s+1} 
    \end{bmatrix}. $ 
Choose $  \alpha_1 (\omega) = \angle \frac{j\omega + 2}{j\omega + 1}, \alpha_2(\omega) = \angle \frac{j\omega+1}{2j\omega+5} $ and $ \alpha_3 (\omega) = \angle \frac{j\omega+2}{j\omega+1}. $ Then one can graphically draw the frequency-wise $\theta$-symmetric SRGs to verify that \eqref{eq:cyclic_system_stability_condition} holds. By Theorem \ref{thm:cyclic_system_theta_SRG}, this implies the feedback  interconnection of a cascade of systems $P_1,P_2,P_3$ is stable.

\section{Conclusions} 
\label{section:conclusion}

In this paper, we study a variant of SRG, termed the $\theta$-symmetric SRG, which integrates gain and refined phase information. Based on this notion, a graphical stability criterion is developed for feedback interconnection of a cascade of systems. Some results in existing literature can be recovered by the over-approximations of $\theta$-symmetric SRG. 

Compared with standard SRG, the $\theta$-symmetric SRG provides a more powerful framework for system characterization, especially when dealing with complex matrices. The frequency-wise $\theta$-symmetric SRG of a system serves as a more suitable MIMO extension of the classical Nyquist plot. This suitability stems not only from its reduced conservatism and exact reduction to the scalar case, but also from its conceptual validity and soundness.

\section*{Acknowledgment}
The authors would like to thank Haoxiang Yang and Prof. Di Zhao of Nanjing University for helpful discussions.

\bibliographystyle{IEEEtran}
\bibliography{TransReference}

\begin{thebibliography}{10}
\providecommand{\url}[1]{#1}
\csname url@samestyle\endcsname
\providecommand{\newblock}{\relax}
\providecommand{\bibinfo}[2]{#2}
\providecommand{\BIBentrySTDinterwordspacing}{\spaceskip=0pt\relax}
\providecommand{\BIBentryALTinterwordstretchfactor}{4}
\providecommand{\BIBentryALTinterwordspacing}{\spaceskip=\fontdimen2\font plus
\BIBentryALTinterwordstretchfactor\fontdimen3\font minus \fontdimen4\font\relax}
\providecommand{\BIBforeignlanguage}[2]{{%
\expandafter\ifx\csname l@#1\endcsname\relax
\typeout{** WARNING: IEEEtran.bst: No hyphenation pattern has been}%
\typeout{** loaded for the language `#1'. Using the pattern for}%
\typeout{** the default language instead.}%
\else
\language=\csname l@#1\endcsname
\fi
#2}}
\providecommand{\BIBdecl}{\relax}
\BIBdecl

\bibitem{Qiu2009IntroductionFeedbackSystems}
L.~Qiu and K.~Zhou, \emph{{Introduction to Feedback Control}}.\hskip 1em plus 0.5em minus 0.4em\relax Upper Saddle River, N.J: Prentice Hall, 2009.

\bibitem{Liu2016Robust_Control}
K.-Z. Liu and Y.~Yao, \emph{{Robust Control: Theory and Applications}}.\hskip 1em plus 0.5em minus 0.4em\relax John Wiley \& Sons, 2016.

\bibitem{MacFarlane1977GeneralizedNyquist}
A.~G.~J. MacFarlane and I.~Postlethwaite, ``The generalized {Nyquist} stability criterion and multivariable root loci,'' \emph{Int. J. Control}, vol.~25, no.~1, pp. 81--127, 1977.

\bibitem{Desoer1980Generalized}
C.~A. Desoer and Y.~T. Wang, ``On the generalized {Nyquist} stability criterion,'' \emph{IEEE Trans. Autom. Control}, vol.~25, no.~2, pp. 187--196, 1980.

\bibitem{Zhou1996Robust}
K.~Zhou, J.~C. Doyle, and K.~Glover, \emph{{Robust and Optimal Control}}.\hskip 1em plus 0.5em minus 0.4em\relax New Jersey: Prentice Hall, 1996.

\bibitem{Chen2024Phase}
W.~Chen, D.~Wang, S.~Z. Khong, and L.~Qiu, ``A phase theory of multi-input multi-output linear time-invariant systems,'' \emph{SIAM J. Control Optim.}, vol.~62, no.~2, pp. 1235--1260, 2024.

\bibitem{Mao2022Phases}
X.~Mao, W.~Chen, and L.~Qiu, ``Phases of discrete-time {LTI} multivariable systems,'' \emph{Automatica}, vol. 142, p. 110311, 2022.

\bibitem{Wang2024First_five_year}
D.~Wang, W.~Chen, and L.~Qiu, ``The first five years of a phase theory for complex systems and networks,'' \emph{IEEE/CAA J. Autom. Sinica}, vol.~11, no.~8, pp. 1728--1743, 2024.

\bibitem{Tits1999Robustness}
A.~L. Tits, V.~Balakrishnan, and L.~Lee, ``Robustness under bounded uncertainty with phase information,'' \emph{IEEE Trans. Autom. Control}, vol.~44, no.~1, pp. 50--65, 1999.

\bibitem{Chen2025SingularAngle}
C.~Chen, D.~Zhao, and S.~Z. Khong, ``The singular angle of nonlinear systems,'' \emph{Automatica}, vol. 181, p. 112515, 2025.

\bibitem{Wielandt1967Topics}
H.~Wielandt, \emph{Topics in the Analytic Theory Matrices}.\hskip 1em plus 0.5em minus 0.4em\relax Lecture Notes, Dept. of Mathematics, Univ. of Wisconsin, Madison, 1967.

\bibitem{Gustafson1968Angle}
K.~Gustafson, ``The angle of an operator and positive operator products,'' \emph{Bull. Amer. Math. Soc.}, vol.~74, pp. 488--492, 1968.

\bibitem{Krein1969AngularLocalization}
M.~G. Krein, ``Angular localization of the spectrum of a multiplicative integral in a {Hilbert} space,'' \emph{Funct. Anal. Appl.}, vol.~3, no.~1, pp. 73--74, 1969.

\bibitem{Gustafson1994Antieigenvalues}
K.~Gustafson, ``Antieigenvalues,'' \emph{Linear Algebra Appl.}, vol. 208, pp. 437--454, 1994.

\bibitem{Chen2025Cyclic}
C.~Chen, W.~Chen, D.~Zhao, J.~Chen, and L.~Qiu, ``A cyclic small phase theorem,'' \emph{IEEE Trans. Autom. Control}, 2025, early access, doi:10.1109/TAC.2025.3617287.

\bibitem{Anderson1973Network}
B.~D.~O. Anderson and S.~Vongpanitlerd, \emph{{Network Analysis and Synthesis: A Modern Systems Theory Approach}}.\hskip 1em plus 0.5em minus 0.4em\relax Upper Saddle River, NJ: Prentice Hall, 1973.

\bibitem{Petersen2010Feedback_control_NI}
I.~R. Petersen and A.~Lanzon, ``Feedback control of negative-imaginary systems,'' \emph{IEEE Control Syst. Mag.}, vol.~30, no.~5, pp. 54--72, 2010.

\bibitem{Hannah2016ScaledRelativeGraph}
R.~Hannah, E.~K.~. Ryu, and W.~Yin, ``Scaled relative graph,'' \emph{UCLA CAM Report}, 2016.

\bibitem{Ryu2022SRG}
E.~K. Ryu, R.~Hannah, and W.~Yin, ``Scaled relative graphs: nonexpansive operators via {2D Euclidean} geometry,'' \emph{Math. Program.}, vol. 194, no.~1, pp. 569--619, 2022.

\bibitem{Pates2021Scaled}
R.~Pates, ``The scaled relative graph of a linear operator,'' \emph{arXiv preprint arXiv:2106.05650}, 2021.

\bibitem{Chaffey2022Rolledoff}
T.~Chaffey, ``A rolled-off passivity theorem,'' \emph{Syst. Control Lett.}, vol. 162, p. 105198, 2022.

\bibitem{Chaffey2023Graphical}
T.~Chaffey, F.~Forni, and R.~Sepulchre, ``Graphical nonlinear system analysis,'' \emph{IEEE Trans. Autom. Control}, vol.~68, no.~10, pp. 6067--6081, 2023.

\bibitem{vandenEijnden2025Phase}
S.~van~den Eijnden, C.~Chen, K.~Scheres, T.~Chaffey, and A.~Lanzon, ``On phase in scaled graphs,'' \emph{arXiv preprint arXiv:2504.21448}, 2025.

\bibitem{Chen2025Graphical}
C.~Chen, T.~Chaffey, and R.~Sepulchre, ``Graphical dominance analysis for linear systems: a frequency-domain approach,'' \emph{arXiv preprint arXiv:2504.14394}, 2025.

\bibitem{Krebbekx2025Graphical}
J.~Krebbekx, R.~T{\'o}th, and A.~Das, ``Graphical analysis of nonlinear multivariable feedback systems,'' \emph{arXiv preprint arXiv:2507.16513}, 2025.

\bibitem{deGroot2025DissipativitySRG}
T.~de~Groot, W.~Heemels, and S.~van~den Eijnden, ``A dissipativity framework for constructing scaled graphs,'' \emph{arXiv preprint arXiv:2507.08411}, 2025.

\bibitem{Baron2025MixedGainPhase}
E.~Baron-Prada, A.~Anta, A.~Padoan, and F.~D{\"o}rfler, ``Mixed small gain and phase theorem: a new view using scale relative graphs,'' \emph{arXiv preprint arXiv:2503.13367}, 2025.

\bibitem{Baron2025StabilityLTI_SRG}
------, ``Stability results for {MIMO LTI} systems via scaled relative graphs,'' \emph{arXiv preprint arXiv:2503.13583}, 2025.

\bibitem{Zhang2025DW_Shell}
D.~Zhang, X.~Yang, A.~Ringh, and L.~Qiu, ``The phantom of {Davis-Wielandt} shell: a unified framework for graphical stability analysis of {MIMO LTI} systems,'' \emph{arXiv preprint arXiv:2507.19918}, 2025.

\bibitem{Tyson1978Dynamics}
J.~J. Tyson and H.~G. Othmer, ``The dynamics of feedback control circuits in biochemical pathways,'' \emph{Prog. Theor. Biol.}, vol.~5, pp. 1--62, 1978.

\bibitem{Hori2011CyclicGene}
Y.~Hori, T.-H. Kim, and S.~Hara, ``Existence criteria of periodic oscillations in cyclic gene regulatory networks,'' \emph{Automatica}, vol.~47, no.~6, pp. 1203--1209, 2011.

\bibitem{Arcak2006DiagonalStability}
M.~Arcak and E.~D. Sontag, ``Diagonal stability of a class of cyclic systems and its connection with the secant criterion,'' \emph{Automatica}, vol.~42, no.~9, pp. 1531--1537, 2006.

\bibitem{Sontag2006SecantCondition}
E.~D. Sontag, ``Passivity gains and the ``secant condition" of stability,'' \emph{Syst. Control Lett.}, vol.~55, no.~3, pp. 177--183, 2006.

\bibitem{Jiang2008Generalization}
Z.~P. Jiang and Y.~Wang, ``A generalization of the nonlinear small-gain theorem for large-scale complex systems,'' in \emph{Proc. 7th World Congr. Intell. Control Autom.}, 2008, pp. 1188--1193.

\bibitem{Pates2023Generalisation}
R.~Pates, ``A generalisation of the secant criterion,'' \emph{IFAC-PapersOnLine}, vol.~56, no.~2, pp. 9141--9146, 2023.

\bibitem{Liang2024Feedback}
J.~Liang, D.~Zhao, and L.~Qiu, ``Feedback stability under mixed gain and phase uncertainty,'' \emph{IEEE Trans. Autom. Control}, vol.~70, no.~2, pp. 1008--1023, 2024.

\bibitem{Li2008Eigenvalues}
C.-K. Li, Y.-T. Poon, and N.-S. Sze, ``Eigenvalues of the sum of matrices from unitary similarity orbits,'' \emph{SIAM J. Matrix Anal. Appl.}, vol.~30, no.~2, pp. 560--581, 2008.

\bibitem{Gustafson1997NumericalRange}
K.~E. Gustafson and D.~K. Rao, \emph{{Numerical Range: The Field of Values of Linear Operators and Matrices}}.\hskip 1em plus 0.5em minus 0.4em\relax Springer, 1997.

\bibitem{Huang2019SRG_Normal}
X.~Huang, E.~K. Ryu, and W.~Yin, ``Scaled relative graph of normal matrices,'' \emph{arXiv preprint arXiv:2001.02061}, 2019.

\bibitem{Paul2015Computation_theta_antieigenvalues}
K.~Paul, G.~Das, and L.~Debnath, ``Computation of antieigenvalues of bounded linear operators via centre of mass,'' \emph{Int. J. Appl. Comput. Math.}, vol.~1, no.~1, pp. 111--119, 2015.

\bibitem{Skogestad2005MultivariableFeedbackControl}
S.~Skogestad and I.~Postlethwaite, \emph{{Multivariable Feedback Control: Analysis and Design}}.\hskip 1em plus 0.5em minus 0.4em\relax John Wiley \& Sons, 2005.

\bibitem{Griggs2012InterconnectionsMixed}
W.~M. Griggs, S.~S.~K. Sajja, B.~D.~O. Anderson, and R.~N. Shorten, ``On interconnections of ``mixed'' systems using classical stability theory,'' \emph{Syst. Control Lett.}, vol.~61, no.~5, pp. 676--682, 2012.

\end{thebibliography}

\end{document}